# Application of the Entropy Production Principle to the Analysis of the Morphological Stability of a Growing Crystal

L. M. Martiouchev (1,2), V. D. Seleznev (1), I. E. Kuznetsova (2)

((1)Department of Molecular Physics, Ural State Technical University, Ekaterinburg, 620002, K-2, RUSSIA; e-mail: martiouchev@mailcity.com

(2) Institute of Industrial Ecology, UB RAS, Ekaterinburg, RUSSIA)

ABSTRACT

Stability of cylindrical and spherical crystals growing from a supersaturated solution (in Mullins-Sekerka's approximation) is considered using the maximum entropy production principle. The concept of the binodal of the nonequilibrium (morphological) phase transition is introduced for interpretation of the obtained results. The limits of the metastable regions are determined. The morphological phase diagrams of stable-unstable growth in the plane (surface energy, supersaturation) are given.

*PACS*: 05.70. Ln- Nonequilibrium thermodynamics, irreversible processes.

*Keywords*: entropy production; pattern formation; morphological transition.

## I. INTRODUCTION

Great attention has been attached recently to the pattern formation during the nonequilibrium growth of crystals owing to its theoretical and practical significance. However, many problems are not solved despite the progress made in this field after the classical works by Ivantsov [1] and Mullins-Sekerka [2]. Let us emphasize just two issues that are directly related to this study:

1. The relationship between the generally accepted analysis for stability (see, e.g., [2,3]) and fundamental principles of the nonequilibrium thermodynamics is not quite understood. Usually these approaches either are opposed one to another or develop independently. Indeed, in accordance with the theoretical studies [4-6], one of the fundamental principles underlying development of a nonequilibrium system is the principle of the maximum entropy production. This principle can be formulated as follows: an arbitrary nonequilibrium system tends to the locally equilibrium state at a maximum rate (with a maximum entropy production). Below it will be shown that in a particular case the said principle reduces to the principle of the maximum possible growth rate of a crystal. However, opinions of researchers concerned with the crystal growth differ widely and change with time as to the principle of the maximum rate. This is explained primarily by the fact that the works [4-6] were overlooked and the principle was formed just intuitively in the literature on physical metallurgy and crystal growth. The prehistory of the maximum rate principle as applied, in particular, to the dendrite growth is briefly as follows. Temkin [7] probably was among those who pioneered the use of the said principle in analytical calculations. The principle served as a criterion for selection of a certain solution from the whole family of possible solutions obtained in terms of a phenomenological model [7]. However, results of the experimental study [8] proved to be considerably different from those of [7], while the theoretical works [9,10], where the stability analysis was used to examine a growing paraboloid in the approximation of the isotropic surface tension, agreed fairly well with the experimental data. As a result, the last theory [9,10], which was called the marginal stability theory, was opposed to the principle of the maximum growth rate and the principle was viewed as invalid. However, we hold to the opinion that differences between the theory based on the maximum rate principle and the experiment might be due primarily to roughness of the phenomenological theory itself, which was used to describe the dendrite growth. In about eight years, some theoretical studies were published indicating to contradictions in the marginal stability theory itself, namely the absence of the steady-state solution of the type representing a needle-like dendrite. As a result, the theory was modified by introducing a weak anisotropy of the surface tension (see reviews [11,12]). The new theory, which is known as the solvability theory, was also based on the stability analysis. The solvability theory stated, among other things, that the solution corresponding only to the maximum growth rate is linearly stable. With development of microscopic solvability theory the problems of dendrite growth seemed to be solved. However, the Hele-Shaw anisotropic experiment and computations in terms of the boundary-layer model revealed one more problem: in the presence of anisotropy, dendrites were not always observed with decreasing undercooling/supersaturation [13-15]. As a result, the dendrite tip split. To overcome this difficulty, it was proposed [13-15] to replace the solvability criterion by a more general principle: the dynamically selected morphology is the fastest-growing one. In other words, if more than one morphology is possible, only the fastest-growing morphology is nonlinearly stable and, hence, observable.

Thus, two approaches to the morphology selection − the stability analysis and the maximum growth rate principle − competed over 30 years of the investigations into the nonequilibrium growth of crystals. Intuitively, both principles are credible. Despite the fact that in some instances the stability analysis suggests the maximum growth rate, it is obvious that in most cases each of the approaches will lead to quantitative, if not qualitative, differences. In our opinion, the search for a more "correct" principle shows no promise. It is necessary to stop opposing these two approaches and try establishing a logic relationship between them.

2. Numerous experimental works and computer simulations show that different morphologies can coexist in a certain region of parameters [13-22]. It is also known from experimental and theoretical studies that if parameters (for example, supersaturation) are changed, transition from one morphology to another may occur as either a jump discontinuity or a discontinuity in the slope of the observed crystal velocity [13-18,23-27]. Therefore the analogy is drawn between phase diagrams and morphology diagrams, and the notions of "the first-order morphology transition" and "the second-order morphology transition" are introduced [13-18,25-28]. The most critical issue in this respect is to find the principle for selection of the probable morphology and construct the complete morphology diagram (with limits of metastable regions) using this principle. But this problem has not been solved yet. A hypothesis was proposed in the literature that the entropy production dominates in the morphology selection far from the equilibrium. However, the appropriate calculations were not made [14-16].

Thus, the above analysis (items 1 and 2) suggests the objective of this study: examine the problem of the morphology selection during nonequilibrium growth of crystals using the principle of the maximum entropy production and, applying the concept of morphology diagrams, show the relationship between this approach and the stability analysis. For convenience and clearness, we shall consider one of the simplest problems: growth of an infinite cylinder and a sphere from the solution in classical Mullins-Sekerka's (MS) approximation [2]. The classical stability analysis of this problem was performed by Mullins, Sekerka [2] and Coriell, Parker [29].

The paper is organized as follows. In Sec.II we analyse the expression of an entropy production for the isotermo-isobaric growth from a solution. The basic local principle about the behaviour of an entropy production is formulated here and all analysis in next sections is a result of this principle. In Sec.III we dwell briefly on the using approximation, and present the calculations of an entropy production for the specific crystal forms. Two cases are considered separately, there are growth of an infinite cylinder and a sphere. A summary and outlook finally is given in Conclusion.

## II. EXPRESSION FOR AN ENTROPY PRODUCTION AT A CRYSTAL GROWTH FROM A SOLUTION

The local entropy production $S$ for the system under study (isothermal-isobaric crystallization, the solvent is fully forced out by the growing crystal) is known to be [30]

$$S = j\,\nabla m, \qquad (1)$$

where $j$ is the flux of the crystallizing component; $\nabla m$ is the chemical potential gradient of the crystallizing component. The expression (1) is applicable to all elements of the volume studied and, in particular, the region near the surface of the growing crystal. In this case, the flux $j$ is

$$j = (C - C_S\,)V. \qquad (2)$$

where $C$ is the constant concentration in the precipitate (crystal density); $V$ is the local growth rate. $C_S$ is the solute concentrations at the crystal surface. Note that the entropy production is proportional to the mass deposition rate of the growing crystal or, in particular, to the interface velocity. It also contains an additional factor equal to the chemical potential gradient. From the results of Ref. [4-6], the following local principle can be formulated for the system under study: If fluctuations in the system have a sufficient amplitude, the condition characterized by the maximum local entropy production is realized. It is worth noting that this principle generalizes the hypotheses of "the fastest growth rate" [13-15,17] and "the largest mass deposition rate" [16]. Corollaries of the maximum entropy production principle for the growing cylindrical and spherical crystals will be analyzed below. This work evolves our previous ideas [31,32], and expands the short publication [33].

## III. THE CHANGE OF AN ENTROPY PRODUCTION OF GROWING CRYSTALS. STABILITY-METASTABILITY-INSTABILITY

### 1. MS approximation

The diffusion controlled growth of a crystal from a supersaturated solution having the initial concentration $C_\infty$ is considered. It is assumed that there is a local equilibrium near each element of the phase interface. Crystallographic factors are disregarded. The diffusion field is described by the Laplace equation, i.e. the condition of the quasistationary diffusion is fulfilled:

$$\left|\frac{C_\infty - C_S}{C - C_S}\right| \le \left|\frac{C_\infty - C_0}{C - C_0}\right| << 1, \qquad (3)$$

where $C_0$ is the solute concentrations at the flat interface. The solution of Laplace equation is carried out with the assumption that the equilibrium concentration on the crystal surface satisfies the equation:

$$C_S = C_0 + C_0\ \Gamma\ K,$$

where $\Gamma$ is a capillary constant (it is proportional surface energy [2,29]), $K$ is the mean curvature.

## 2. The growth of cylindrical crystal

In MS approximation the behavior of an infinitesimal distortion of the particle interface of the form $F(\varphi,z)=cos(k\varphi)cos(k_z z/R)$ ($k$ being a positive integer and $k_z$ having any real positive value) is analyzed. The equation of the distorted cylinder surface is

$$r\ (\varphi,z,t)= R(t) + \delta(t)F(\varphi,z), \qquad \delta(t)<<R(t) \tag{4}$$

where $R$ is the unperturbed cylinder radius; $\delta$ is the perturbation amplitude; $t$ is the time. Using the general solution of the Laplace equation for a slightly deformed cylinder and neglecting all powers of $\delta$ greater than unity, we have [29]

$$V=\overset{\bullet}{R}+\overset{\bullet}{\delta}\ F(\varphi,z), \tag{5}$$

$$\overset{\bullet}{\delta}(t) = \frac{C_0 D}{(C-C_R)R^2}\frac{1}{A_\varphi}\left\{\frac{C_\infty - C_0}{C_0}(H-1)-\frac{\Gamma A_\varphi}{R}\left[H(k^2+k_z^2-1)+\frac{H-1}{A_\varphi}\right]\right\}\delta, \tag{6}$$

where $D$ is the diffusivity; $H = H(k,k_z) = k_z[K_{k-1}(k_z)+K_{k+1}(k_z)]/2K_k(k_z)$, $K_k(k_z)$ is a modified Hankel function; $C_R=C_0(1+\Gamma/R)$; $A_\varphi= -0.5\ \ln(\eta^2/\varphi^2)$; $\varphi$ is found from $\varphi^2\ln(\eta^2/\varphi^2) + (C_s-C_R)/(C-C_R) = 0$; $\ln(\eta^2)$ is Euler's constant; $\overset{\bullet}{R}\equiv dR/dt$; $\overset{\bullet}{\delta}\equiv d\delta/dt$.

From (6) it follows that the cylinder growth is stabilized by the surface energy and is destabilized by the concentration gradient. The expression (6) also suggests that perturbations increase if the cylinder radius exceeds the critical value ($R_C^S$) [29]:

$$R_C^S = \{1+[HA_\varphi(H-1)^{-1}][k^2+k_z^2-1]\}\ R_C^*, \tag{7}$$

where $R_C^*=\Gamma C_0/(C_\infty-C_0)$ is the critical radius of the nucleation theory. Note that the ratio $R_C^S/R_C^*$ is not a constant (if $k$ and $k_z$ are constant) but depends on supersaturation (because $A_\varphi$ is a function of supersaturation).

The formulas (6)-(7) completely determine stability of a growing cylindrical particle with respect to an infinitely small perturbation [29].

Now let us analyze this problem in terms of the thermodynamic approach.

Using (1) and (2), determine the difference between the entropy productions for the perturbed ($S_p$) and unperturbed ($S_n$) growth of an infinite cylinder (we find the entropy production of a volume element, which is located near the crystal surface and has the area cut out by the solid angle $dW$ and the unit thickness). For definiteness, we shall use the ideal solution approximation, i.e. $\bar{N}m \sim \bar{N}C/C$. As a result,

$$DS_C \equiv S_p - S_n = (S_p r - S_n R)\, dW \sim \{(C-C_S)^2 V^2\, r/C_S - (C-C_R)^2\, \dot{R}^2 R/C_R\}\, dW.$$

If we take into account (3)-(6) and the fact that the steady growth rate of a cylinder from a solution is determined by the formula $\dot{R} = D(C_\infty - C_R)/[\,RA_i\,(C-C_R)\,]$ [29], then

$$DS_C \sim \left\{ 2H - 1 - [\,2A_i\,H(k^2 + k_z^2 - 1) + 2H - 1\,]\frac{R_C^*}{R} - (k^2 + k_z^2 - 1)\frac{C_\infty - C_R}{C_R}\frac{R_C^*}{R}\right\} d F(j,z) dW. \quad (8)$$

If the relative supersaturation is assumed to be small, the expression (8) becomes

$$DS_C \sim (2R\,\dot{d} + d\,\dot{R})dW \sim \left\{ 2H - 1 - [\,2A_i\,H(k^2 + k_z^2 - 1) + 2H - 1\,]\frac{R_C^*}{R}\right\} d F(j,z) dW. \quad (9)$$

Choose the direction $(j,z)$ corresponding to the maximum value of $F(j,z)$. The most dangerous condition (with respect to the cylinder growth disturbance) is realized for this direction. Using (9), it is possible to show that within the interval $[R_C^*, R_C^S]$ where the cylinder radius can change, the function $DS_C$ is positive at $R > R_C^b$:

$$R_C^b = \left[ 1 + \frac{2A_i\,H(k^2 + k_z^2 - 1)}{2H - 1}\right] R_C^*. \quad (10)$$

So, the difference between the entropy productions for the perturbed and unperturbed cylinder growth changes sign at the point $R_C^b$, which differs from $R = R_C^S$ obtained in terms of the MS theory. This discrepancy ($R_C^b \uparrow R_C^S$) is caused by the second term in (9), which is proportional to the perturbation amplitude $d$ (the entropy production of the considered volume element depends not only on the linear growth rate of the crystal but also on the change in its surface area). When $R = R_C^S$, in accordance with the MS theory, the crystal loses stability with respect to an infinitesimal perturbation. In terms of the equilibrium thermodynamics, this point can be called the spinodal of the nonequilibrium phase transition. In the interval $[R_C^b, R_C^S]$, according to MS results, the growing cylinder is stable with respect to infinitely small perturbations. However, in this interval the entropy production of the volume element in the presence of a perturbation is larger than in the absence of a perturbation. Therefore, from the maximum entropy production principle it follows that the growth of the distorted cylinder is preferable. This contradiction disappears, if one assumes that the cylinder growth is metastable, i.e. unstable with respect to some small, but finite perturbations. Let us refer to the region $[R_C{}^b, R_C{}^S]$ as metastable and the point $R_C{}^b$ as the binodal of morphological transition.

The tangential solute fluxes arising near the cylinder surface is the physical reason of the metastable behavior. When $R > R_C{}^b$, these fluxes go towards the forward bulge in the cylinder surface (i.e., where $F(l,z)$ is a maximum), because the solute is distributed nonuniformly at the solid angles during the growth of the surface perturbation.

The entropy production of the volume element under consideration is shown in Fig. 1.

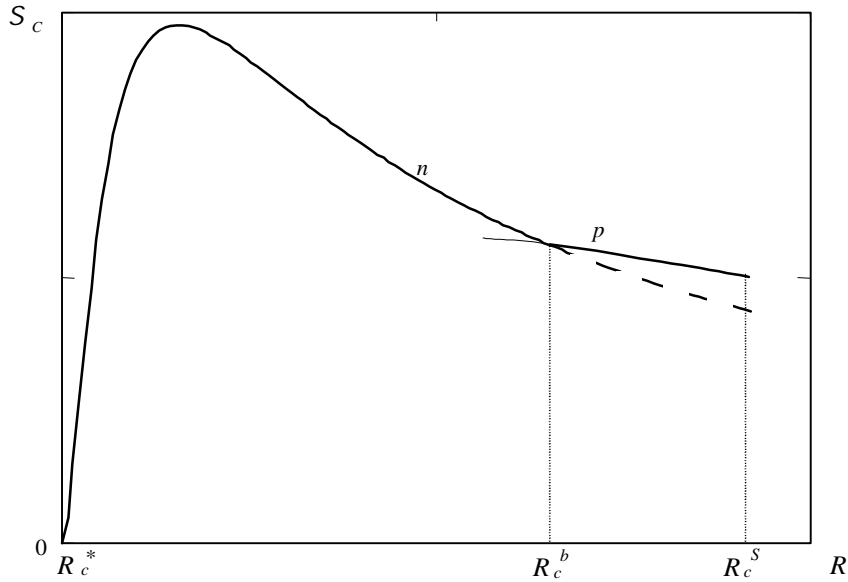

Fig. 1. Local entropy production $S_C$ of the volume element under consideration as a function of the cylinder radius $R$. The metastable region is shown with the dashed line.
n - unperturbed growth; p - perturbed growth (with the forward bulge on the cylindrical surface).

The formulas (7) and (10) readily suggest that the metastable interval $R_C{}^S - R_C{}^b = R_C{}^* A \, H(k^2 + k_z{}^2 - 1)/((H-1)(2H-1))$ exists at any physically possible parameters. This interval expands with increasing surface energy and shrinks with increasing relative supersaturation $(C_y - C_0)/C_0$.

The morphological phase diagram of stable-unstable growth plotted in the plane (supersaturation, capillary constant) is presented in Fig. 2.

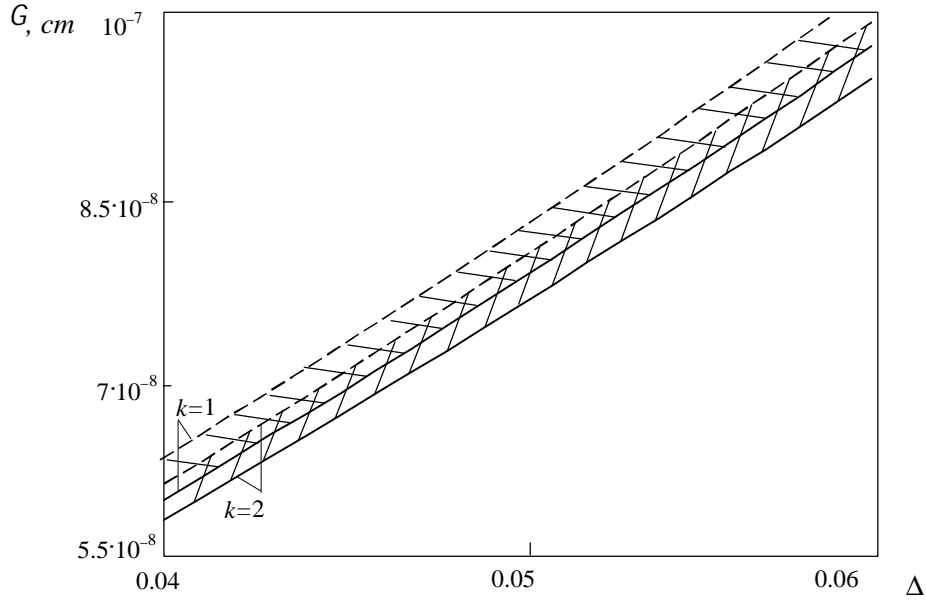

Fig. 2. Morphological phase diagram of stable-unstable growth of cylindrical crystal plotted in the parameter space of the capillary constant $G$ versus the relative supersaturation $\Delta=(C_v\text{-}C_0)/C_0$ for $k=1,2$ and $k_z=10$. The solid curves denote the spinodals and the dashed curves stand for the binodals. The metastable region is shaded. The stable growth is above the binodal and the unstable growth is under the spinodal. The curves are plotted for $R=5\cdot10^{-4}$cm, $C=6C_0$.

This diagram was constructed using the formulas (7) and (10) at a fixed value of the radius. The diagram is convenient because in real experiments researchers always work with a specific substance and a specific space scale ("fixed microscope magnification"), and, varying parameters (for example, supersaturation), observe either morphology (nonequilibrium phase). The regions of existence and coexistence of various morphologies are shown in this diagram.

Figure 3 gives $R_C^b$ and $R_C^S$ in units of $R_C^*$ as a function of $k_z$ ($k=1,2,3$). One can see from this figure that binodals and spinodals of different perturbing harmonics may intersect one another. Let us analyze the evolution of this system with reference to the diagram (Fig.3). Consider the crystal growth in the presence of some perturbation on the $z$-axis (for example, $k_z=6.5$) and three perturbations of $j$ ($k=1,\ k=2,\ k=3$). In other words, the straight line AC describes the evolution of the cylindrical crystal. The growth is stable up to the point A. The crystal growth becomes metastable with respect to perturbations with $k=1$ if the cylinder radius is within the interval AB. In this interval the loss of stability is possible only if fluctuations have a sufficient amplitude. If several cylindrical crystals grow under these conditions, part of them continue the cylindrical growth, while the other lose stability. Obviously, the quantity of stable cylindrical crystals decreases with increasing $R$.

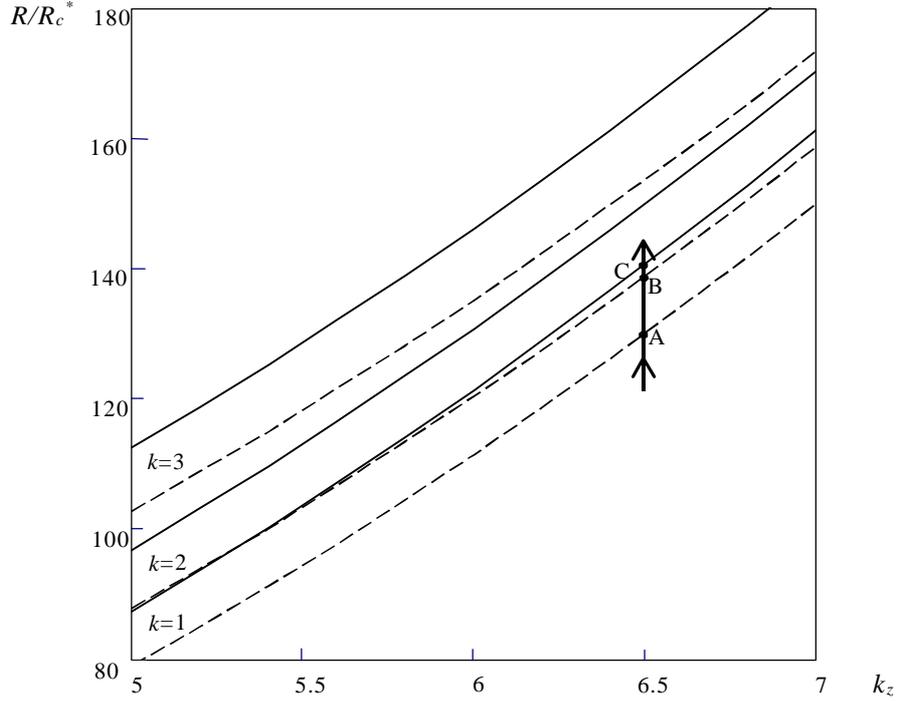

Fig. 3. $R_c^b$ and $R_c^S$ in units of $R_c^*$ as a function of $k_z$ for $k$=1,2,3 at $A_1$=2.9 (corresponding to $(C_s-C_0)/C_0$=0.05). The solid curves are spinodals $R_c^S$ and the dashed curves are binodals $R_c^b$. For details see text.

Starting from the point B, development of perturbations with both $k$=1 and $k$=2 is possible. In this case, if several crystals grow simultaneously, most of the crystals lose stability at the perturbation with $k$=1, some also lose stability but at the perturbation with $k$=2, and few crystals preserve the cylindrical shape in the interval BC. Starting from the point C, all cylindrical crystals lose stability with respect to an infinitesimal perturbation with $k$=1. From (7), (10) and Fig. 3 it is seen that an unlimited number of morphological phases may coexist if high-frequency perturbations occur along the $z$-axis.

Following the lines of reasoning adopted elsewhere [29] and using (10), consider the case when long-wave perturbations occur in the $z$-direction only ($k$=0, $k_z \le 1$).

1. If $k_z$=1, then $R_C^b = R_C^S = R_C^*$ and the crystal is unstable from the initial moment of growth.

2. If $0.6 \le k_z < 1$, then $R_C^S < R_C^b < R_C^*$ and the cylindrical surface is always unstable. This result is similar to the result obtained in terms of the theory of morphological stability [29]. It is explained by the fact that both the surface energy and the concentration gradient are conducive to development of perturbation.

3. In accordance with [29], the case when $k_z < 0.6$ is most unusual: The concentration gradient in (6) changes sign and, thus, favors the decrease in perturbations. But the surface energy promotes the growth of perturbations. As a result, the cylindrical surface is unstable at $R < R_C^S$ and is stable at $R>R_C{}^S$ [29]. Our calculations also revealed the presence of a metastable interval $(R_C{}^S<R<R_C{}^b)$ separating stable and unstable regions.

For some special perturbations ($k=1$ and $k_z<1$) the interval of possible variation of the cylinder radius $[R_C{}^*, R_C{}^S]$ may contain the maximum of the function $\Delta S_C$ at $R_C{}^{ex} \cong 1.5 R_C{}^b$. Therefore, when the crystal grows from $R_C{}^b$ to $R_C{}^S$, instability first increases and then drops. In this particular case the reentrant behavior takes place.

One more interesting case of perturbations is realized at $k=0$ and $k_z \gg 1$. In this case $H(k,k_z) \cong k_z+0.5$ [29] and $R_C{}^S \to R_C{}^b$. Thus, the metastable region vanishes.

## 3. The growth of spherical crystal

In MS approximation, the behaviour of a infinitesimal distortion of the particle form, described by a single spherical harmonic $Y_{lm}$, is investigated. The equation of a perturbed sphere surface is

$$r(q,j,t) = R(t) + d(t)Y_{lm}(q,j), \qquad d(t) \ll R(t) \tag{11}$$

where $R$ is the nonperturbated sphere radius. Using a common solution of the Laplace equation for a slightly deformed particle, one can determine (all powers of $d$ greater than unity is neglected) [2]:

$$V = \dot{R} + \dot{d}\, Y_{lm}, \tag{12}$$

$$\dot{d}(t) = \frac{C_0 D(l-1)}{(C-C_R)R^2}\left\{\frac{C_\infty - C_0}{C_0} - \frac{G}{R}[(l+1)(l+2)+2\,]\right\}d, \tag{13}$$

where $C_R = C_0(1+2G/R)$.

From (13) it follows, that $l$ harmonic will increase if the radius of the sphere is more than critical ($R_S{}^S$):

$$R_S{}^S = R_S{}^*[(l+1)(l+2)/2+1], \tag{14}$$

where $R_S{}^* = 2GC_0/(C_\infty - C_0)$ is the critical nucleation radius.

The obtained formulas (13), (14) completely define the stability of a growing spherical particle with respect to infinitely small perturbation [2].

Using (1) and (2), we find the difference between the entropy productions of volume element (with area, cut out solid angle $dW$, and with unit thickness) near the crystal surface for the cases of the perturbed ($S_p$) and nonperturbed ($S_n$) spherical particle growth in MS approximation. For distinctness we'll use approximation of a ideal solution again. As result

$$DS_S \equiv S_p - S_n = (S_p \, r^2 - S_n \, R^2) \, dW \sim \{(C-C_S)^2 V^2 \, r^2/C_S - (C-C_R)^2 \, \overset{\bullet}{R}{}^2 R^2/C_R\} \, dW. \tag{15}$$

If we take into account (3), (11)-(13) and the fact that the steady growth velocity of a sphere from a solution is determined by the formula $\overset{\bullet}{R} = D(C_\infty - C_R)/[R(C-C_R)]$ [2] then

$$DS_S \sim \left\{ 1 - \left[ \frac{(l+1)(l+2)+2}{2}(l-1)+1 \right] \frac{R_S^*}{R} - \left[ \frac{(l-1)(l+2)}{4} \frac{C_\infty - C_R}{C_R} \right] \frac{R_S^*}{R} \right\} \partial Y_{lm} dW.$$

In assuming of smallness of relative supersaturation the difference between the entropy productions of the perturbed and nonperturbed cases is

$$DS_S \sim (R \, \overset{\bullet}{\partial} + \partial \, \overset{\bullet}{R}) dW \sim \left\{ 1 - \left[ \frac{(l+1)(l+2)+2}{2}(l-1)+1 \right] \frac{R_S^*}{R} \right\} \partial Y_{lm} dW \tag{16}$$

We choose direction $(\partial, j)$, which corresponds to the maximum value of $Y_{lm}$ as in the previous case. Using (16) it is possible to establish that on the $[R_S^*, R_S^S]$ interval the function $DS_S$ for considered direction is positive at $R > R_S^b$:

$$R_S^b = R_S^* \frac{l^3 + 2l^2 + l - 2}{2l}. \tag{17}$$

So, the difference between the entropy productions in cases of the perturbed and nonperturbed sphere growth changes the sign in the point $R_S^b$, which differ from $R=R_S^S$. And so, all conclusions (about binodal, metastable region and coexistence of morphological phase) are similar to ones obtained above for the infinite cylinder. There is only one speciality: the metastable regions for different perturbed harmonics don't cross themselves in sphere case. This feature is shown on the Fig.4 (compare with Fig.2).

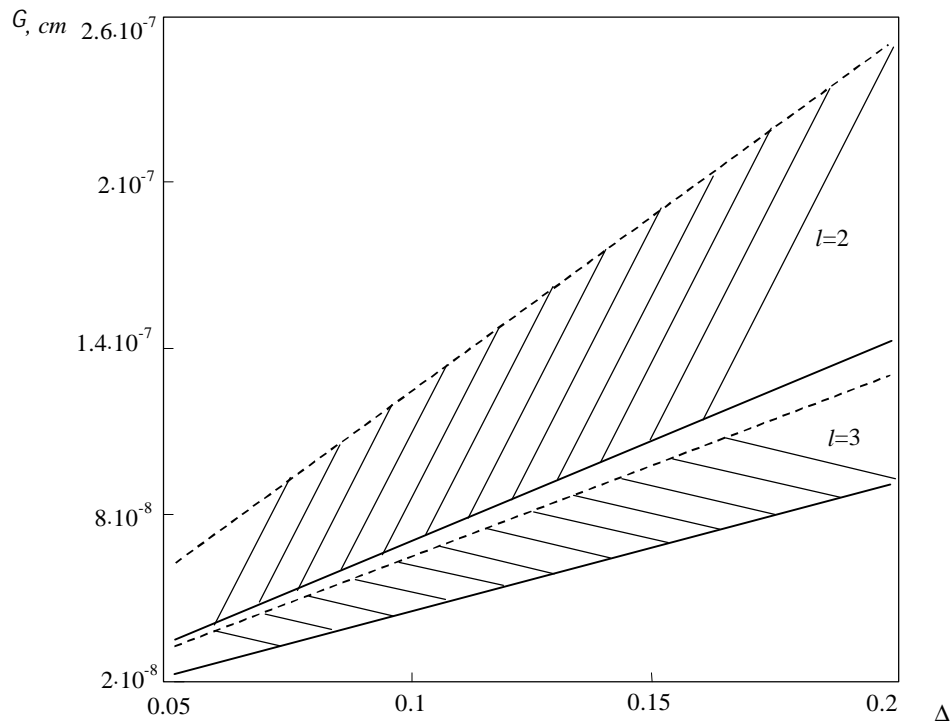

Fig. 4. Morphological phase diagram of stable-unstable growth of spherical crystal plotted in the parameter space of the capillary constant $G$ versus the relative supersaturation $\Delta=(C_s-C_0)/C_0$ for $l=2,3$. The solid curves denote the spinodals and the dashed curves stand for the binodals. The metastable region is shaded. The stable growth is above the binodal and the unstable growth is under the spinodal. The curves are plotted for $R=10^{-5}$cm.

## CONCLUSION

So, the study into the morphology stability of growing cylindrical and spherical crystals allows proposing the relationship between the thermodynamic approach, which uses the maximum entropy production principle, and the stability analysis. Maximum principle is not an alternative, but a supplement to the theory of the stability analysis used traditionally, and contributes to solving of the morphological transition's problem.

It was shown that the loss of stability of a cylindrical and spherical crystals during their growth from a supersaturated solution is the first-order nonequilibrium phase transition (presence of a metastable region and a sharp increase in the entropy production at $R>R^b$). Calculations were made in MS approximation for ideal solution and small supersaturations. The experimental verification of the limits of the stable growth ($R^b$) is complicated, because these sizes ($R^b$, $R^S$) and ($R^S-R^b$), are of the order of $R^*$ (i.e. about 1 μm). From (7),(10) and (14),(17) it follows that in such experiments one needs to provide small supersaturation, use a system (crystal-solution) with a large surface tension and, if possible, eliminate low-frequency perturbations. Nonlinear analysis can be used for theoretical testing of the problems under consideration, but it is enough difficult and cumbersome. Note, that weakly nonlinear morphological stability analysis for the cylinder and the sphere [34,35] allows to calculate nonlinear critical radius which is situated in the metastable region found in our work.

The possibility of the coexistence of the numerous morphological phases is the main difference between the results of cylindrical and spherical growth problems. However, as it seems to us, it is not a fact of principle, but it follows from using approximation; under other conditions (nonideal solution, any supersaturations) several morphological phases can coexist at a spherical crystal growth.

It is worth noting that papers describing coexistence of different morphologies are numerous, but all of them deal either with experimental or computer simulation methods [13-22]. A merit of the proposed method is the possibility of the analytical determination of the morphology diagrams (with stable, unstable and metastable regions).